# Domain Wall-Magnetic Tunnel Junction Spin Orbit Torque Devices and Circuits for In-Memory Computing

Mahshid Alamdar[1*], Thomas Leonard[1*], Can Cui[1], Bishweshwor P. Rimal[1], Lin Xue[2], Otitoaleke G. Akinola[1], T. Patrick Xiao[3], Joseph S. Friedman[4], Christopher H. Bennett[3], Matthew J. Marinella[3], and Jean Anne C. Incorvia[1]

[1]Electrical and Computer Engineering Dept., University of Texas at Austin, Austin TX USA, email: incorvia@austin.utexas.edu

[2]Applied Materials, Santa Clara CA USA,   [3]Sandia National Laboratories, Albuquerque NM USA,   [4]Electrical and Computer Engineering Dept., University of Texas at Dallas, Richardson TX USA   *Co-first author

There are pressing problems with traditional computing, especially for accomplishing data-intensive and real-time tasks, that motivate the development of in-memory computing devices to both store information and perform computation[1]. Magnetic tunnel junction (MTJ) memory elements can be used for computation by manipulating a domain wall (DW), a transition region between magnetic domains. Three leading device types that use MTJs and DWs for in-memory computing are majority logic[2–4], mLogic[5–9], and DW-MTJs[10–14]. But, these devices have suffered from challenges: spin transfer torque (STT) switching of a DW requires high current, and the multiple etch steps needed to create an MTJ pillar on top of a DW track has led to reduced tunnel magnetoresistance (*TMR*)[15-16]. These issues have limited experimental study of devices and circuits. Here, we study prototypes of three-terminal domain wall-magnetic tunnel junction (DW-MTJ) in-memory computing devices that can address data processing bottlenecks and resolve these challenges by using perpendicular magnetic anisotropy (PMA), spin-orbit torque (SOT) switching, and an optimized lithography process to produce average device tunnel magnetoresistance *TMR* = 164%, resistance-area product *RA* = 31 $\Omega - \mu m^2$, close to the *RA* of the unpatterned film, and lower switching current density compared to using spin transfer torque. A two-device circuit shows bit propagation between devices. Device initialization variation in switching voltage is shown to be curtailed to 7% by controlling the DW initial position, which we show corresponds to 96% accuracy in a DW-MTJ full adder simulation. These results make strides in using MTJs and DWs for in-memory and neuromorphic computing applications.

Computing today faces walls when processing data-intensive and unstructured tasks. Memory access in modern computers can dominate as much as 96% of computing time[17]. SRAM idle leakage can consume over 20% of the total power of a computation[18,19]. For internet of things applications, there is a large bottleneck in analog to digital conversion, where analog readout can consume over 70% of the active power[20]. The development of in-memory computing elements could alleviate the memory wall between computation and memory on-chip, as well as between devices and the cloud by allowing more computation to be performed on the device with a subset of information going to the cloud. The nonvolatility of such elements would reduce idle leakage.

The DW-MTJ is a candidate in-memory computing device, with previous prototypes showing fanout and concatenation in circuits[11]. Simulation work has shown a single DW-MTJ can perform a NAND function, the devices can be cascaded to build a one-bit adder[10], and, recently, a 32-bit adder with registers was simulated entirely from DW-MTJs[14]. Versions of DW-MTJ devices have also found application in neuromorphic computing circuits for artificial intelligence[21–24]. If the challenges of these devices could be resolved, they could remove the bottleneck between computation and memory.

Figure 1a is a schematic of the studied DW-MTJ device. A heavy metal/ferromagnet/oxide trilayer, e.g. Ta/CoFeB/MgO, is patterned into a DW track with an output MTJ centered on top of the track. Voltage $V$ applied between the *IN* and *CLK* terminals moves the position of a DW in the DW track. The DW position determines the centered MTJ resistance and therefore the output current from the *OUT* terminal to the next DW-MTJ device in the circuit[10,11]. The MTJ resistance $R_{MTJ} = R_P$ when the magnetic layers are parallel (P) and $R_{MTJ} = R_{AP}$ when antiparallel (AP). $Oe_1$ to $Oe_2$ is an Oersted field electrode used to nucleate the initial DW position[25].

The MTJ stack was grown by Applied Materials in an Endura Clover™ physical vapor deposition system[26] with layers Si(substrate) /SiO$_2$(100) /Ta(10) /CoFeB(1.2) /MgO(1) /CoFeB(1.9) /[Co/Pt](5) /Ru(0.9) /[Co/Pt](6.9) /Ta(1) /Ru(3); numbers are in nm and brackets represent multilayers. Using a physical property measurement system, the average film properties were TMR = 168 ± 6% and resistance-area product $RA$ = 35 ± 2 $\Omega - \mu m^2$.

The film stack was patterned using electron beam lithography and ion beam etching (see Methods). Scanning electron microscope images of the devices are shown in Fig. 1b-c for DW track widths $w$ = 350-450 nm. To facilitate keeping the DW from exiting the device, notches are fabricated on both sides of the track.

The out-of-plane major and minor hysteresis loops of the unpatterned film, Fig. 2a-b, show PMA as well as an offset minor loop, showing field coupling between the free and pinned layers. Figure 2c-d shows $R_{MTJ}$ vs. out-of-plane magnetic field $H(\hat{z})$ for four different devices under test. The device properties are summarized in Table 1 compared to the unpatterned film. The average device TMR = 164 ± 17%, close to the unpatterned average TMR = 168 ± 6%, but with increased standard deviation. The device average $RA$ = 31 ± 3 $\Omega - \mu m^2$, close to the unpatterned average $RA$ = 35 ± 2 $\Omega - \mu m^2$. These results show both the TMR and RA is maintained after patterning, and that we have high TMR close to the expected highest TMR (~200%) seen in PMA CoFeB-MgO MTJs[26]. This is a large improvement over previous results where the patterning degraded the TMR to 10-15%[15,16]. High TMR is important in MTJs for logic applications, since it determines the current separation between the 0 and 1 states.

The average switching field of the patterned free layer in the major loop (Fig. 2c) $H_{c,free}$ = -118 ± 5 mT, showing the free layer switching has increased from the unpatterned $H_{c,free}$ = 5 mT. The average switching field of the pinned layer in the major loop $H_{c,pinned}$ = -175 ± 21 mT, close to the unpatterned film (190 mT). This shows that the interfacial anisotropy of the pinned layer is stronger than the shape anisotropy of patterning it into a 350-450 nm diameter pillar; co nversely, the free layer anisotropy is increased when it is patterned into a narrow wire shape.

In the minor loop, Fig. 2d, the average AP to P switch $H_{c,AP-P} = 1.5$ mT $\pm$ 1.7 mT and average $H_{c,P-AP} = -118$ mT $\pm$ 5 mT. The field offset shows the coupling between the free layer DW track and pillar-structure pinned layer is still present, requiring use of an in-plane applied magnetic field during device testing to overcome this coupling; this field could be removed in the future by further optimizing the film stack.

We study the DW-MTJ switching behavior vs. $V$ when an external magnetic field is used to set the initial magnetization. First, a saturation field $H_{SAT}(\hat{z}) = 50$ mT is applied to set all magnetization out of plane. Then, $H_B(\hat{x})$ is applied, which serves the purpose of 1) overcoming the inter-layer coupling field, 2) tilting the magnetization away from $\hat{z}$ for SOT switching, and 3) perhaps nucleating a DW. $V$ is then applied between the *IN* and *CLK* terminals to switch the free layer magnetization. Figure 2e-f shows example $R_{MTJ}$ vs. $V$ for 5 cycles for two of the devices, with the field conditions applied between cycles to re-initialize the DW. A cycle-to-cycle distribution of switching voltage $V_C$ is observed. There are two switching ranges which can be attributed to two locations in the DW track where the DW tends to be nucleated. It is clear that $V_C$ depends on the energy landscape of the DW track and the initial DW location, showing a varying but not fully random switching voltage distribution. Converting $V_C$ to current density $J_C = \frac{V_C}{R_w A_w}$, where $R_w$ is the measured resistance between *IN* and *CLK* and $A_w$ is the cross-sectional area of the patterned Ta/CoFeB wire, results in $J_C = 1.0$-$4.8 \times 10^{11}$ A/m$^2$, reduced from STT prototypes with $J_C \approx 2 \times 10^{12}$ A/m$^2$ [15].

A two-device buffer circuit is set up by wire bonding *OUT* of Device 1 to *IN* of Device 2 (Fig. 3a). After field saturation at -125 mT, $H_B = -1$ mT is applied, setting both AP. Then $V = 2$ V is applied between *IN1* and *CLK1* to write Device 1 P with Device 2 remaining AP. To then read Device 1 while writing Device 2, a voltage pulse is applied between *CLK1* and *CLK2*. Figure 3b shows, for 10 re-initialization cycles, that the current through Device 1 switches $R_{MTJ2}$ of Device 2. $R_{MTJ1} = R_P$ for all 10 cycles, showing Device 1's logic state is stable against being read. Repeated testing of the devices over multiple weeks shows an increase in $R_P$ and $R_{AP}$, and therefore a decrease in *TMR*, than the initial voltage switching of Fig. 2. We attribute this to wear that could be improved by using shorter pulses and not operating the devices above $V_C$.

When a magnetic field is used to initialize the DW position, there is high initialization variation in $V_C$ of the devices. We see binning of $V_C$ in different voltage groups, evidence that the field is nucleating a DW in one of a few different locations in the DW track, with $V_C$ depending on the energy landscape of that location. Thus, we expect improved cycle-to-cycle variation if the DW is initialized in the same position more repeatably. After applying $H_{SAT} = -120$ mT to bring Device 2 to the AP state, 50 ns voltage pulses are applied from $V_{Oe} = 0$-$2$ V through the additional Oe field line electrode ($Oe_1$ to $Oe_2$) that is centered on top of the left notch in the DW track. Over 20 cycles, $V_{Oe} = 2$ V has 100% probability of switching $R_{MTJ}$ from $R_{AP}$ to $R_P$. After applying $V_{Oe} = 2$ V, $V = 0$-$5$ V is applied from *IN* to *CLK* with $H_B = -100$ mT, and this is repeated from field saturation for 10 cycles. Figure 4a shows the cycle-to-cycle initialization variation using external field nucleation, compared to Fig. 4b when current nucleation is used with the Oersted line. Defining percent variation $var = \frac{1}{2} * \frac{V_{C,max} - V_{C,min}}{V_{C,avg}}$, $var$ decreases from 105% in Fig. 4a to 7% in Fig. 4b. This is evidence that the Oe field line introduces a DW at or near the left notch in the device, which is underneath the Oe field line. Since $V_{Oe}$ switches $R_{MTJ}$, most likely a 180° DW is nucleated that changes the magnetization under the MTJ. Thus, in Fig. 4 the resistance switch is opposite in the two plots. The average depinning voltage increases from 2 V to 3.5 V, which agrees with the hypothesis that a DW is nucleated at the notch with a lithographically-designed pinning location that will increase $V_C$ at the expense of better controlling the DW position, important for device design considerations. In an operating circuit, it is expected the DW will not have to be often nucleated since once created it will be pushed back and forth, with nucleation used to refresh the data when needed.

Device success depends on viability in a functional circuit. We have simulated a DW-MTJ 32-bit adder; see Ref[14] for details. In the simulation, parameter variation is increased from 0 to 25% and the full adder accuracy (% correct data output) is calculated, with results shown in Fig. 4c. The measured $var$ = 7% corresponds to an adder accuracy of 96%. Further optimization of the device to $var$ = 4% would produce almost no errors.

In conclusion, DW-MTJs are a main class of magnetic random access memory-like devices for in-memory computing and artificial intelligence applications. This work overcomes major challenges in DW-MTJ devices by showing high *TMR* and maintained *RA* after patterning, measuring lowered switching current density using SOT, characterizing the initialization variation in switching behavior, showing bit propagation in a two-device circuit, and showing that current initialization of the DW position can lead to low enough cycle-to-cycle variation for circuit applications. The work provides design parameters for DW-MTJ circuits and for improved future device and circuit design, and it motivates film growth and device optimization to remove the need for external fields.

ACKNOWLEDGMENT

The authors acknowledge funding from Sandia National Laboratories (SNL) Laboratory Directed Research and Development. SNL is managed and operated by NTESS under DOE NNSA contract DE-NA0003525. The work was done at the Texas Nanofabrication Facility supported by NSF grant NNCI-1542159 and at the Texas Materials Institute (TMI).

METHODS

Patterning of all features was performed with a Raith electron beam lithography (EBL) tool using Ma-N 2405 negative-tone resist and PMMA-A4 positive-tone resist; etching of the features was done using an AJA International ion miller. All processing steps were performed below 170 °C to prevent heating damage to the PMA layers. Device testing was carried out using a home-built testing setup with the device wire bonded to a chip carrier and placed near a tunable electromagnet.

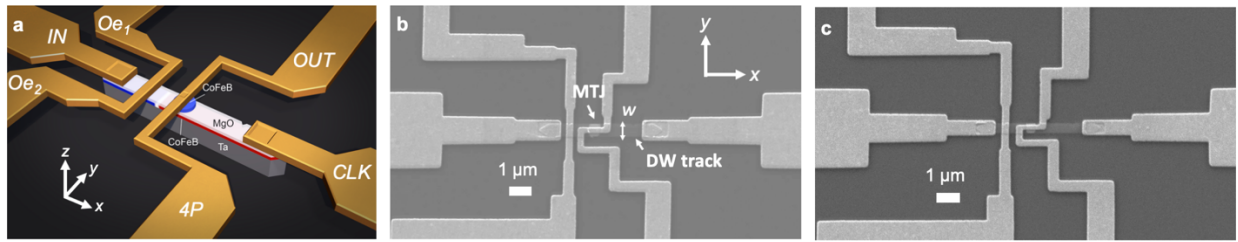

**Fig. 1 | DW-MTJ device. a**, Schematic of the three-terminal DW-MTJ device with *IN*, *CLK*, and *OUT* terminals, Ta/CoFeB/MgO DW track, and output MTJ (blue circle). Red and blue represent domains in ±z. Only essential layers are shown and not to scale. The *4P* terminal is used to measure the four-point resistance of the MTJ; $Oe_1$ to $Oe_2$ is an Oersted field electrode. **b**, Top-down scanning electron microscope image of *w* = 450 nm and **c**, *w* = 350 nm device prototypes with patterned DW track and output MTJ labeled.

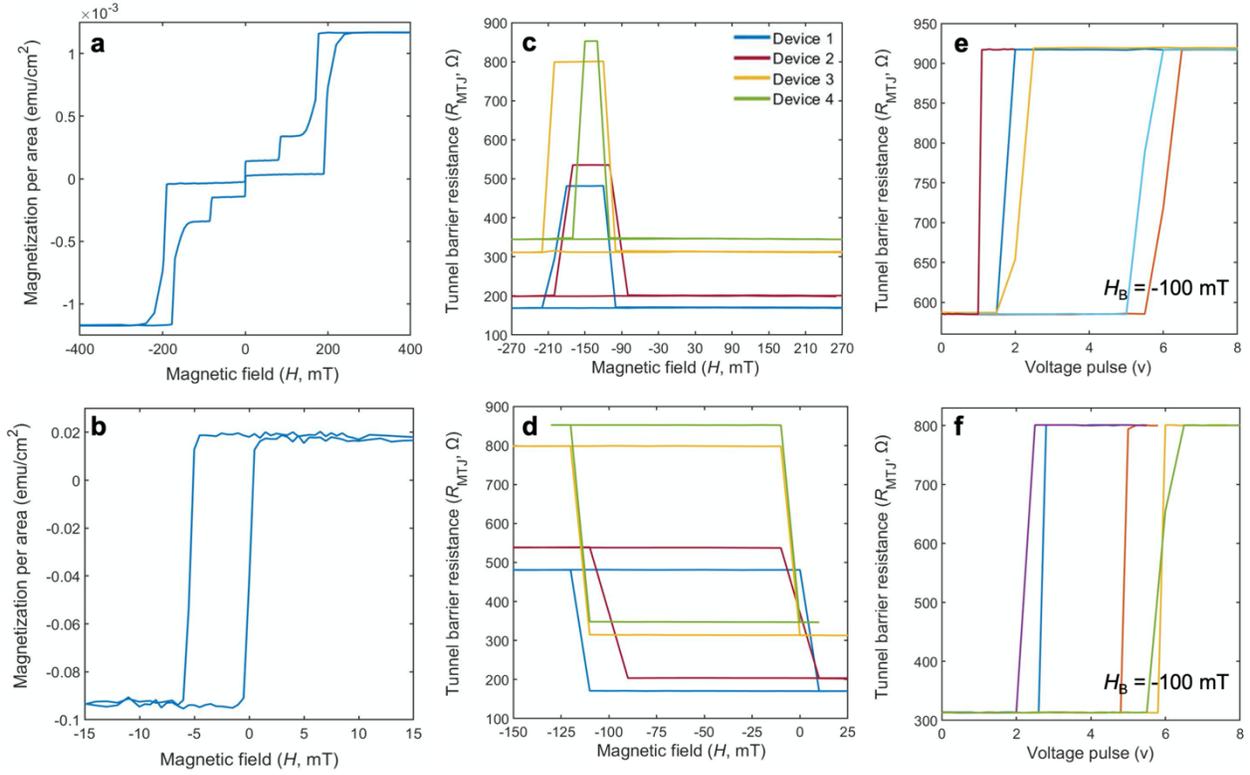

**Fig. 2 | Film and device switching behavior. a**, Unpatterned film out-of-plane major field loop and **b**, minor field loop. **c**, $R_{MTJ}$ vs. out-of-plane magnetic field for half of the major loop and **d**, for the minor loop where only the free layer is switched, for the four devices in Table 1. **e**, Example $R_{MTJ}$ vs. $V$ (1 μs pulses of increasing amplitude) between *IN* and *CLK* device terminals for 5 cycles for Device 2 and **f**, for Device 3.

| # | w (nm) | TMR (%) | RA (Ω-μm$^2$) | $R_P$ (Ω) | $R_{AP}$ (Ω) | $|H_C|$ AP → P (mT) | $|H_C|$ P → AP (mT) | $|H_C|$ Hard Layer (mT) |
|---|---|---|---|---|---|---|---|---|
| Unpatterned Film | - | 168 ± 6 | 35 ± 2 | - | - | 0.5 | 5 | 190 |
| Device 1 | 450 | 185 | 27 | 169 | 481 | 4 | 120 | 220 |
| Device 2 | 450 | 168 | 32 | 199 | 534 | 1 | 110 | 170 |
| Device 3 | 350 | 155 | 30 | 313 | 800 | 0.5 | 120 | 200 |
| Device 4 | 350 | 146 | 33 | 347 | 855 | 0.5 | 120 | 150 |
| Devices Avg. | - | 164 ± 17 | 31 ± 3 | - | - | 1.5 ± 1.7 | 118 ± 5 | 175 ± 21 |

**Table 1 | Device properties.** Switching behavior of the four example devices compared to the unpatterned film.

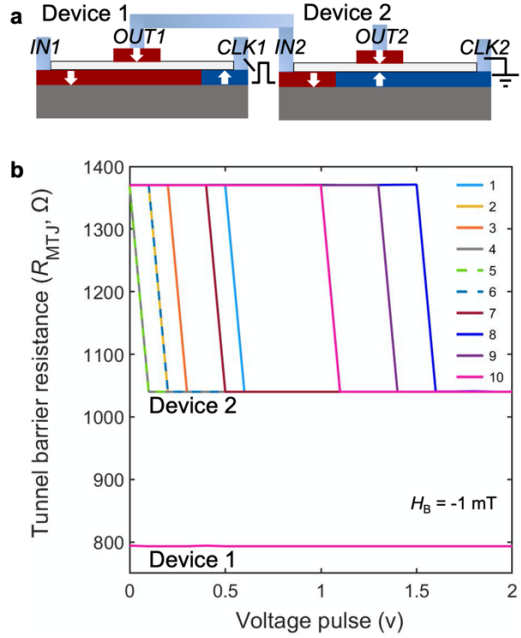

**Fig. 3 | Two-device circuit. a**, Cartoon of buffer circuit just before *V* is applied to read Device 1 and write Device 2, with *OUT* of Device 1 connected to *IN* of Device 2. **b**, MTJ resistance of both devices measured over 10 operation cycles; Device 1 resistance is stable with only cycle 10 (pink) visible.

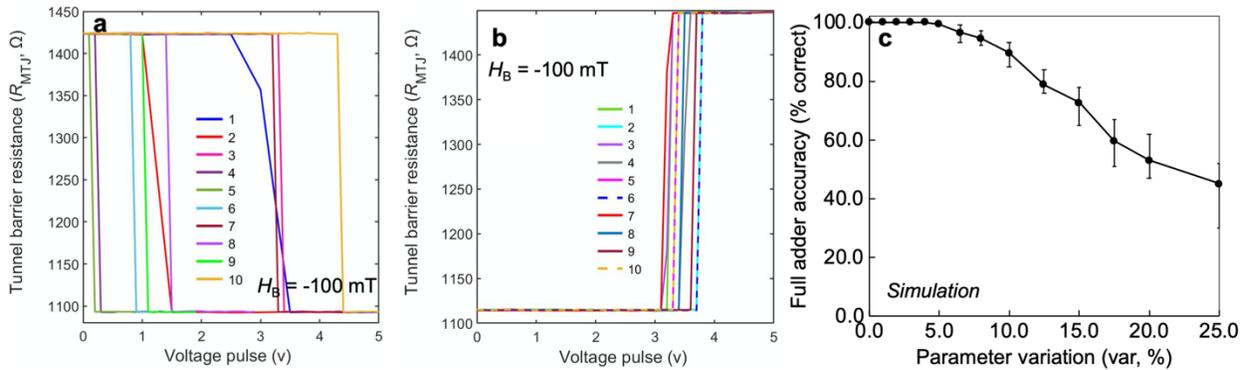

**Fig. 4 | Initialization variation. a**, Device 2 10 cycle variation when the DW is initialized with external magnetic field. **b**, Device 2 10 cycle variation when the DW is initialized using the Oersted field line. **c**, Simulated full adder accuracy vs. device parameter variation.